\documentclass[11pt,twoside]{article}
\usepackage{newpasp}

\usepackage{epsf}

\index{Poggianti, B.M.}
\index{van Gorkom, J.}

\begin{document}

\title{Environmental effects on Gas and Galaxy Evolution in Clusters}
\author{Bianca M. Poggianti}
\affil{Oss. Astron., vicolo dell'Osservatorio 5, 35122 Padova, I}
\author{Jacqueline van Gorkom}
\affil{Dep. of Astron., Columbia University, New York, NY 10027}

\begin{abstract}
The cluster environment is the site of important transformations 
in galaxies at a relatively recent cosmological epoch: the galactic 
morphological types and star formation activity have evolved significantly 
during the last few Gyrs. How is this related to the galactic gas content? 
The first part of this paper briefly presents the optical observations
of galaxies in distant clusters, while 
the second part focuses on the link between the \hbox{H {\sc i}} content 
and the star formation activity, 
presenting the results of a multiwavelength study of the cluster
A2670 (z=0.08) where 42 galaxies have been detected in \hbox{H {\sc i}}. 
We show the star formation properties of the \hbox{H {\sc i}}-normal and \hbox{H {\sc i}}-deficient 
galaxies and discuss their dependence on the location within the cluster.
Finally, we compare A2670 with
clusters at intermediate redshift and with the Coma cluster.
\end{abstract}
\vspace{-0.5in}
\section{Galaxy evolution in clusters at optical wavelengths}
Galaxy evolution in clusters has some interesting characteristics:

a) it is {\em rapid}, because significant evolution is observed
already between $z\sim 0.2-0.3$ and $z=0$;

b) it mainly involves disk galaxies. 
As we will discuss, most of the action is taking place
in spiral and S0 galaxies while not much happens to the bulk of the ellipticals
except passive evolution, at least up to $z\sim 0.6$.

c) from the optical point of view two types of galaxy evolution are observed:
a transformation of the galactic morphologies (i.e. Hubble types) and a change
in the star formation activity, therefore the stellar population content.

Most of the distant cluster work presented in this section is a result of the
MORPHS collaboration (involving W. Couch, A. Dressler, R. Ellis, 
A. Oemler, B.M. Poggianti and I. Smail)
that studied 10 rich clusters at $z=0.4-0.5$ using WFPC2-HST imaging
and ground-based optical spectroscopy (Smail et al. 1997, Dressler et 
al. 1999 [D99]).

\subsection{Morphological evolution}
The relative frequency of the various Hubble types evolves with redshift. 
Roughly speaking, spirals are a factor of 2 to 3 more abundant at $z\sim 0.5$
than at $z=0$, while S0 galaxies are proportionally less abundant in the 
distant clusters (Dressler et al. 1997 [D97]). 
Hence, a large number of the S0 galaxies we observe in clusters today seem
to have evolved from the numerous spirals observed at $z=0.5$ (see also
Fasano et al. 2000 [F00]).
In contrast, the fraction of ellipticals at $z=0.5$ is already as large 
or larger than in the nearby universe (D97). 
Furthermore, many of the spirals have 
disturbed morphologies, in some cases due to an obvious interaction/merger.

The morphology-density relation changes with redshift as well:
at low z it was found to
be valid {\em in all types of clusters} (Dressler 1980), while at higher 
redshift a qualitatively similar relation
exists only in centrally concentrated clusters and is absent
in low-concentration ones (D97, F00).

\subsection{Stellar population evolution}
The first evidence for strong evolution of the stellar populations 
came from photometric data
and was uncovered by Butcher \& Oemler (1984), who found
an excess of blue galaxies in clusters at $z \ge 0.2$. 
The large number of spectra now available has allowed to study the
star formation histories of the cluster galaxies in greater detail
(Poggianti et al. 1999 [P99]).

\begin{table}
{\scriptsize
\caption{\scriptsize 
Spectral classification scheme and fraction of galaxies as 
a function of spectral class. This
is purely a {\em spectral} classification which gives no information
about the {\em morphology} of the galaxy.}
\begin{tabular}{lcccccc}
\tableline
Type & $\rm H\delta$ & [O{\sc ii}] & Comments & Distant & Distant & A2670 \\
     &     \AA          &  \AA           &    & clusters & field  &       \\
\tableline
\noindent Non star-for.: &&&&&&\\
k      & $<3$  &  absent  &  passive, elliptical-like & 0.48$\pm$0.04 & 0.30$\pm$0.07 & 0.73$\pm$0.08 \\
k+a    & $>3$  &  absent  &  post-SB/post-starfor. & 0.21$\pm$0.02 & 0.06$\pm$0.03 & 0.08$\pm$0.03 \\
&&&=E+A&&&\\
Spiral-like: &&&&&&\\
k(e)   & $<4$  &  weak    & early-spiral like & see e(c) & see e(c) & see e(c)\\
e(c)   & $<4$  & moderate & spiral-like & 0.14$\pm$0.02 & 0.29$\pm$0.06 & 0.16$\pm$0.04 \\
&&&&&&\\
Starbursting: &&&&&&\\
e(a)   & $>4$  & weak-mod. & dusty SB & 0.11$\pm$0.02 & 0.11$\pm$0.04 & 0.00\\
e(b)   & any &  $>$40 & SB w. strong emi. lines & 0.05$\pm$0.01 & 0.16$\pm$0.05 & 0.02$\pm$0.01 \\
\tableline
\tableline
\end{tabular}
}
\end{table}

Galaxies in distant clusters show a large variety of spectral characteristics
(D99, see Table~1), namely:

-- totally passive spectra without signs of current or recent star formation,
that resemble those of nearby ellipticals (k-type). 

-- post-starburst/post-starforming galaxies with no ongoing star formation,
that were forming stars at some time during the last Gyr. These are known
as ``E+A'' galaxies, and will be called ``k+a'''s in this paper;

-- spectra of star-forming galaxies that resemble those of normal, quiescent 
nearby spirals (e(c) type);

-- starburst galaxies (e(b) and e(a) types). 
The e(a) spectra  are characterized by a peculiar combination of spectroscopic
features (Table~1) and have been proposed to be spectra of 
starburst galaxies that are highly obscured by dust 
(P99, Poggianti \& Wu 2000). It is important to note
that {\em by definition} the difference between e(a) and k+a galaxies is
the presence of the [OII] emission. Neverthless, it is possible that some
of the k+a's are extreme e(a) galaxies in which the [OII] line is totally 
extincted by dust (Smail et al. 1999, see also Owen's contribution in these
proceedings).

The fraction of galaxies as a function of the spectral class are given
for the MORPHS clusters and field in Table~1 (P99).
In clusters at $z=0.4-0.5$, 
we observe a large fraction of k+a galaxies (21\%
of the total cluster population) and 
globally a much higher proportion (30\%) of 
star-forming galaxies than at $z=0$, of which about 1/3 are dusty
starbursts (10\%). 
Post-starburst galaxies are much more frequent in the distant
cluster environment 
(21\%) than in the distant field (6\%), while 
the proportion of star forming galaxies is 
higher in the field and dusty starburst spectra are 
present in both environments with a similar incidence (11\%).

Comparing the morphological and spectral classifications, 
the majority of Es and S0s have stellar populations that are old,
passively evolving (k type), while most of the post-starburst and of 
the star-forming spectra belong to spiral galaxies. Overall, the star
formation properties of the spirals are quite surprising: only 10\% of them
have spectra similar to nearby field spirals, about
20\% are starbursting and the majority
($\sim 70$\%) are not currently forming stars, being either k+a or k types.


These results are consistent with field spirals/groups infalling into the
clusters and having their star formation rate truncated by some mechanism
related to the cluster environment, and then turning into S0s {\em at 
some later time}.
The fact that most of the k+a spectra belong to galaxies that are still 
classified as spirals 
indicates that the process responsible for halting the star formation
must not affect the morphology, or at least
not on the same timescale. A process acting only on the gas content of
an infalling galaxy -- such as an ICM-ISM interaction -- is therefore a good
candidate for producing k+a spiral galaxies
(see Bower's and Vollmer's contributions in these proceedings).

Finally, it is important to note that -- although the most obvious effect
of the cluster environment is the halting of the star formation (k+a galaxies)
-- these results do not exclude that at the same time
the cluster could provoke an enhancement of SF in some subset of galaxies.

\section{A multiwavelength study of Abell 2670}
The physical mechanism, or mechanisms,
responsible for the observed evolution still need to be identified. 
The \hbox{H {\sc i}} information can be very valuable in discriminating among
the candidate processes because it can provide the link between the
evolution of the morphologies/star formation and the gas content
and can aid reconstructing the building history of the cluster.

We and a number of other collaborators are carrying out an \hbox{H {\sc i}}+optical
study of 20 clusters out to z=0.2; here we present the results
of the first of these clusters, Abell 2670, obtained in collaboration with
K. Dwarakanath, R. Guhathakurta, A. Shambrook, R. Sharples.
Abell 2670 is a rich cluster at z=0.08 which at first sight appears
a relaxed, dynamically evolved cluster from the optical and the X-ray data.
However, the cD galaxy has a large offset from the cluster mean velocity and 
there is evidence for substructure from a statistical analysis of the 
redshift catalog which shows the presence of at least 2 and probably 3
subsystems merging almost along the line of sight (Bird 1994).

Our group has acquired VLA \hbox{H {\sc i}} imaging, 
CTIO UBVRI wide-field imaging and Keck images of selected areas 
of A2670 and has analyzed $\sim 300$ optical spectra
from Sharples et al. (1988) and ROSAT archive images (Hobbs \& Willmore 1997).
The VLA imaging consists of 3 pointings,, mapping the entire volume of 
the cluster and covering the whole velocity space of cluster members
over an area 5 Mpc by side down to a limit of $10^8 \, M_{\odot}$ of \hbox{H {\sc i}}.
The VLA data also provide as a by-product radio continuum information.

We begin by comparing in Table~1 the star formation properties of A2670 
with those of the MORPHS clusters for galaxies down to
the same absolute magnitude.
The two main differences are that k+a galaxies are much less abundant and
that e(a) spectra are missing among the luminous galaxies of A2670.

\begin{figure}[t!]
\vbox {
  \begin{minipage}[l]{1.0\textwidth}
   \hbox{
       \begin{minipage}[l]{0.8\textwidth}
       {\centering \leavevmode \epsfxsize=\textwidth 
        \hspace{-0.8in} \epsfbox{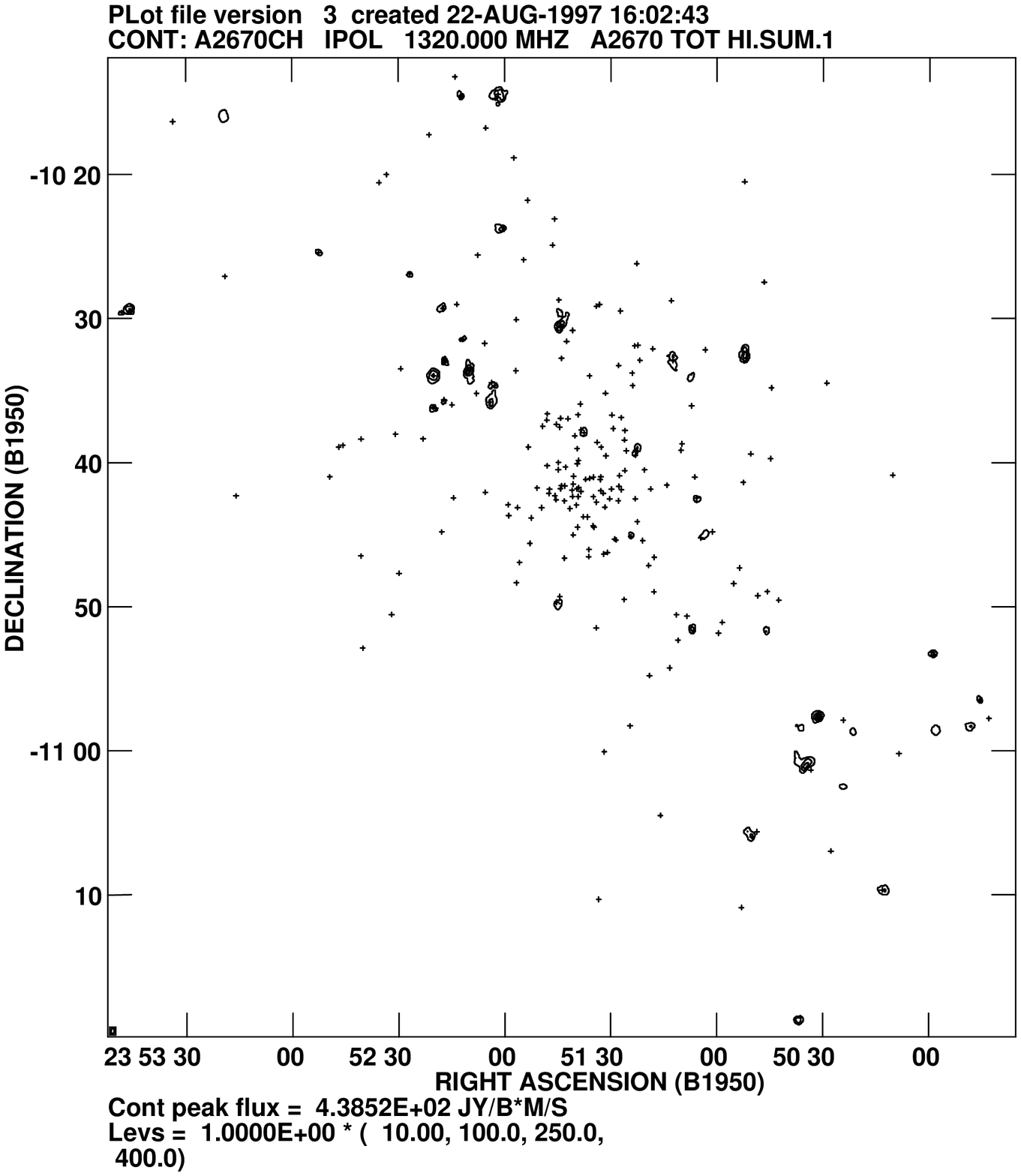}}
       \end{minipage} \  \hfill \
       \begin{minipage}[r]{0.5\textwidth}
       {\centering \leavevmode \epsfxsize=\textwidth 
        \hspace{-1.in} \epsfbox{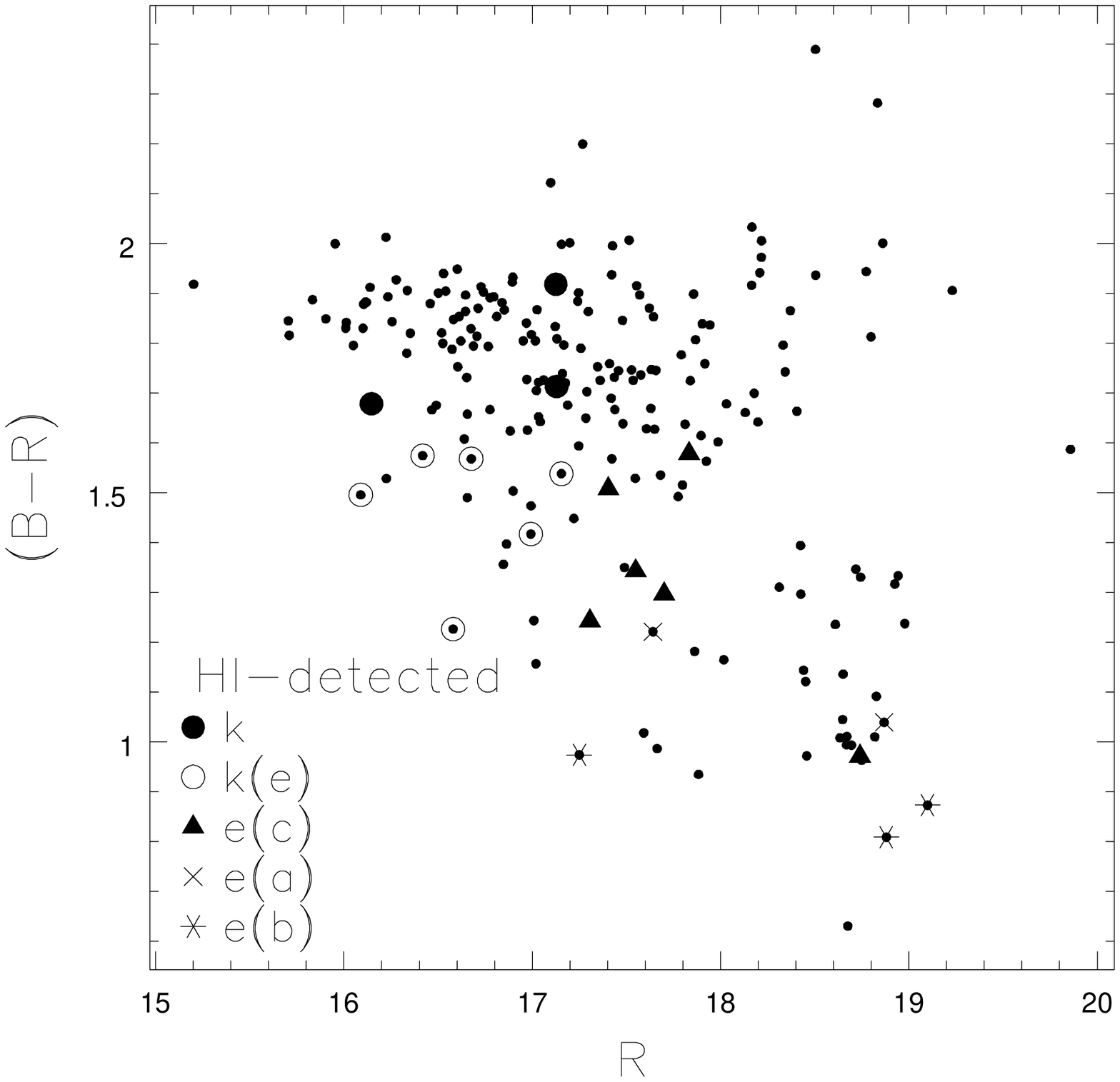}}
       \end{minipage} \  \hfill \
  }
  \end{minipage} \  \hfill \
  \begin{minipage}[l]{1.0\textwidth}
\caption{\small Left: Total \hbox{H {\sc i}} image of A2670. Contours indicate 
\hbox{H {\sc i}} and crosses are spectroscopically confirmed cluster members.
Right: Color-magnitude diagram of galaxies members of A2670.}
  \end{minipage}
}
\label{fig2}
\vspace*{-0.4cm}
\end{figure}

The total \hbox{H {\sc i}} image is presented in Fig.~1 (left panel), where
the 42 \hbox{H {\sc i}} detections are shown as contours.
A group of very rich \hbox{H {\sc i}} galaxies with \hbox{H {\sc i}} masses up to
more than $10^{10} \, M_{\odot}$ is visible in the north-east region. 
This group has a small velocity dispersion
and its mean velocity is higher than the cluster mean and close to the velocity
of the cD galaxy. This appears to be a subsystem that hasn't gone through the 
cluster centre yet. There are somewhat sparser \hbox{H {\sc i}} detection SW and there are 
no \hbox{H {\sc i}} detections within the central 250 kpc. This lack of detections in the
core is not
due to the morphology-density relation, because in this region of the
cluster there are at least 5 bright
spirals of types Sb or later that fall within the \hbox{H {\sc i}} velocity range and yet 
are undetected. 

\begin{figure}[t!]
\vbox {
  \begin{minipage}[l]{1.0\textwidth}
   \hbox{
       \begin{minipage}[l]{0.5\textwidth}
       {\centering \leavevmode \epsfxsize=\textwidth 
        \hspace{-0.4in} \epsfbox{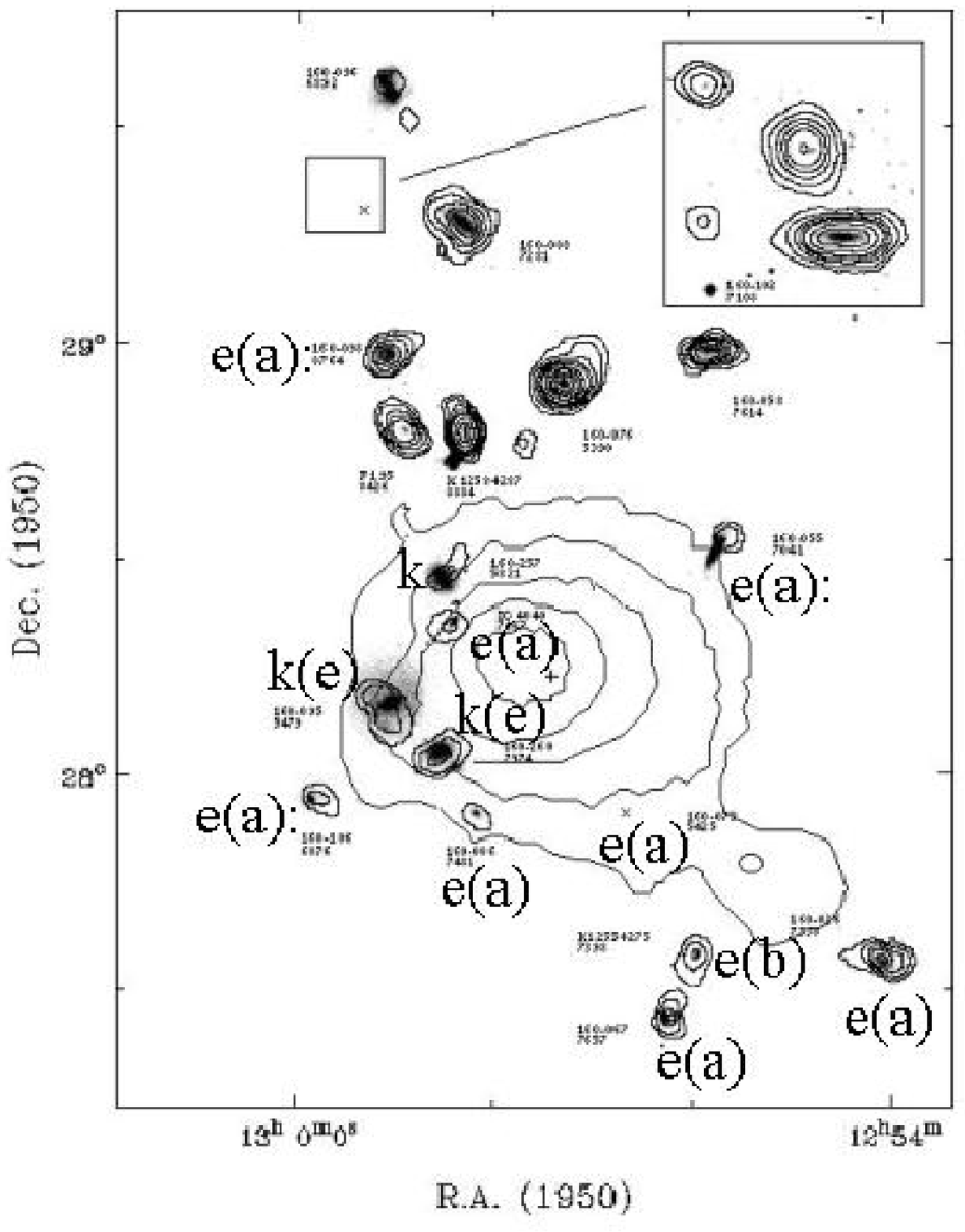}}
       \end{minipage} \  \hfill \
       \begin{minipage}[r]{0.5\textwidth}
       {\centering \leavevmode \epsfxsize=\textwidth 
        \hspace{0.1in} \epsfbox{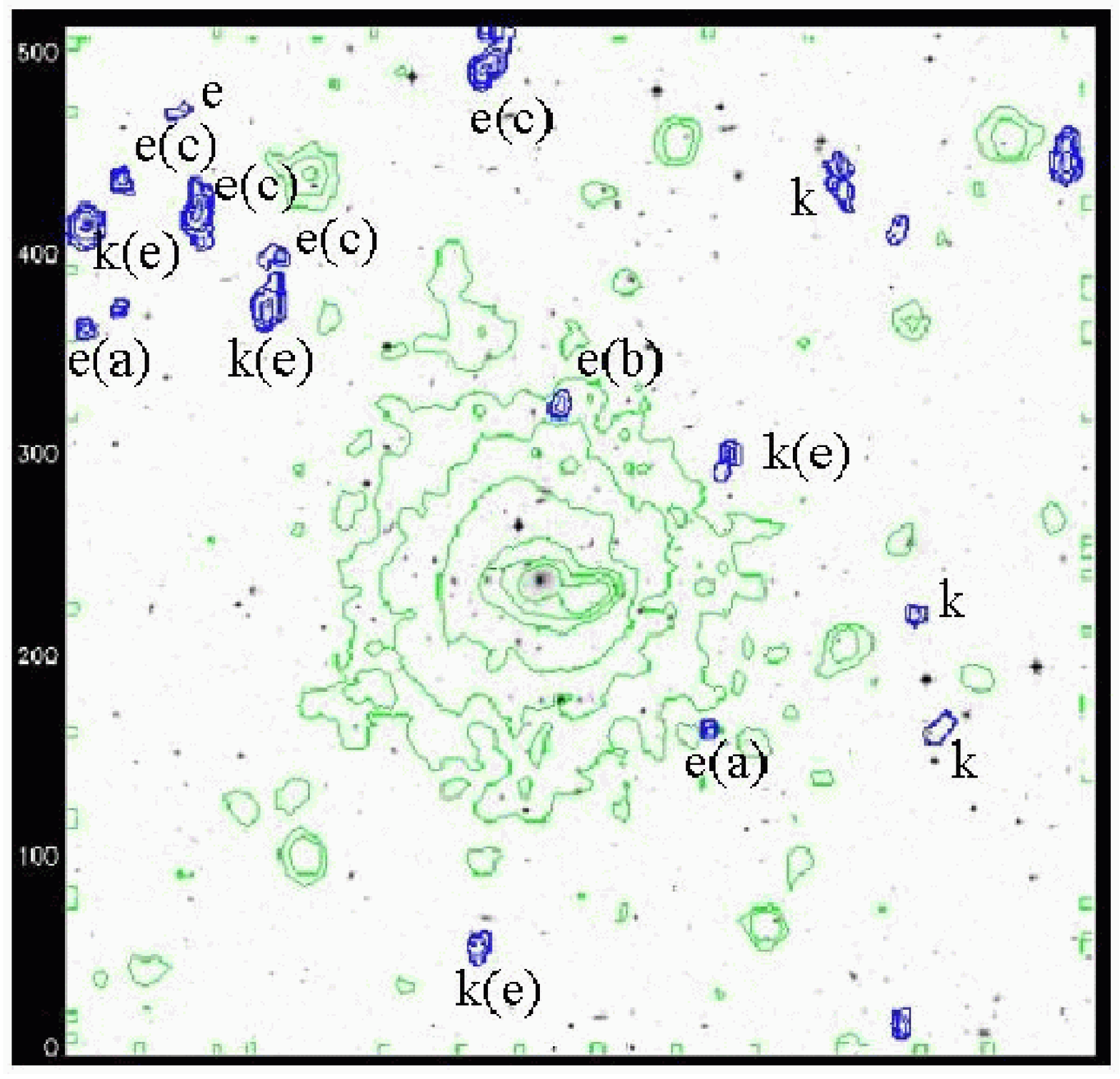}}
       \end{minipage} \  \hfill \
  }
  \end{minipage} \  \hfill \
  \begin{minipage}[l]{1.0\textwidth}
\caption{\small 
Left: Composite plot of individual \hbox{H {\sc i}} maps of spiral galaxies
in Coma (Bravo-Alfaro et al. 2000, see contribution in these proceedings).
Right: \hbox{H {\sc i}} image of the central region of A2670. 
Dark contours indicate \hbox{H {\sc i}} and lighter
contours show the ROSAT X-ray emission.}
  \end{minipage}
}
\label{fig2}
\vspace*{-0.4cm}
\end{figure}

The \hbox{H {\sc i}} image of the central $\sim 2.7 \, \rm Mpc^2$ is presented in Fig.~2
(right panel)
together with the ROSAT contours and the spectral types of the \hbox{H {\sc i}} galaxies.
The six galaxies in projection closest to the cluster centre have asymmetric
\hbox{H {\sc i}} distributions and in some cases the \hbox{H {\sc i}} is displaced from the optical. 
Their
\hbox{H {\sc i}} masses are in the range $4 \times 10^8 - 10^9 \, M_{\odot}$. The star
formation properties of these galaxies show a bimodal behaviour: 2 of them are
starbursts (optically faint e(b) and e(a)) 
while the other 4 are rather passive (k and k(e)
types). None of them has a vigorous, spiral-like star formation (e(c) type).
In contrast, 
the \hbox{H {\sc i}}-rich group at NE comprises many e(c) and k(e) types,
as one would expect for a group of field galaxies composed of early and late 
type spirals.

It is worth noticing the existence of
a population of k-type, yet gas rich galaxies.
Their spectral classification is confirmed by their red colors as shown in
Fig.~1 (right panel). 
This figure shows that in general there is a good correlation between
the spectral type and the color, with the k, k(e), e(c), e(a) and e(b) types
becoming progressively bluer.

Among the \hbox{H {\sc i}} detections there aren't any secure k+a
galaxies and this is consistent with the post-starburst/post-starforming
galaxies being devoid of significant amounts of neutral gas.
K+a galaxies are also undetected in radio-continuum: out of the 23 
cluster members with radio continuum emission, 17 have spiral-like star 
formation (e(c)'s) and 6 are totally passive (k's).

It is interesting to contrast the results of A2670 with those of 
the Coma cluster, in which 
Bravo-Alfaro et al. (2000) have carried out a detailed \hbox{H {\sc i}} imaging
study of 19 spirals (see also Bravo-Alfaro et al. in these proceedings)
finding that the \hbox{H {\sc i}} properties of the cluster galaxies
are strongly correlated with the projected distance from the cluster centre.
The most strongly \hbox{H {\sc i}}-deficient galaxies are located roughly within the extent
of the central X-ray emission and show asymmetries in the \hbox{H {\sc i}} distribution 
and often a shift between the optical and the \hbox{H {\sc i}} position.
Surprisingly, in Coma most of the \hbox{H {\sc i}}-detected galaxies with 
disturbed \hbox{H {\sc i}} morphology -- which are likely to be currently
interacting with the ICM -- have e(a) spectra (Fig.~2, left panel). 
Those that are not dusty
starbursts are rather passive (1 k and 2 k(e) types).

The \hbox{H {\sc i}} and optical data together are consistent with clusters
accreting groups of galaxies that have both their gas content and
their SF properties altered by the impact with the X-ray IGM.
\hbox{H {\sc i}}-detected galaxies in the outskirts of A2670 -- those undisturbed
in their \hbox{H {\sc i}} content and morphology -- have SF properties typical of
field spirals, while both in A2670 and in Coma
those galaxies that still retain some gas -- 
but appear to be already affected by the cluster environment in their \hbox{H {\sc i}}
content -- are either passive or starbursting with significant amounts of dust.
The relative proportions in the two clusters are different: Coma 
has numerous starbursts among the \hbox{H {\sc i}}-disturbed galaxies
while A2670 has mostly passive spectra, as if
Coma has been ``caught in the act'' and A2670 is seen after/before an
infalling group begins to be affected by the ICM.
This is suggestive of a SF enhancement due to the
interaction with the IC gas in Coma.

Here we have presented an \hbox{H {\sc i}}-optical comparison
of two low-z clusters.
In the nearby universe we are observing the tail of evolutionary
processes that are prominent in clusters at slightly higher redshift,
hence it will be extremely interesting to study the \hbox{H {\sc i}} content in more distant
clusters, at z=0.2 and beyond, where the populations of infalling 
star-forming galaxies are a significant fraction of the global cluster
population.

\end{document}